\documentclass[letterpaper,aps,prl,twocolumn,showcaps,superscriptaddress,showpacs,amsmath,amssymb]{revtex4}
\usepackage{graphicx}
\usepackage{color}

\newcommand{\mean}[1]{\langle #1\rangle}
\newcommand{\unit}[1]{\,\mathrm{#1}}

\renewcommand{\vec}[1]{\mathbf #1}

\newcommand{\eps}{\varepsilon}
\newcommand{\lam}{\lambda}
\newcommand{\sig}{\sigma}

\newcommand{\vhi}{\varphi}
\newcommand{\x}{\vec r}
\newcommand{\nois}{\boldsymbol\xi}
\newcommand{\Dr}{D_\text{r}}

\newcommand{\kT}{k_\text{B}T}
\newcommand{\Pe}{\text{Pe}}

\begin{document}

\title{Dynamical clustering and phase separation \\ in suspensions of
  self-propelled colloidal particles}

\author{Ivo Buttinoni}
\affiliation{II. Institut f\"ur Physik, Universit\"at Stuttgart,
  D-70569 Stuttgart, Germany}
\author{Julian Bialk\'e}
\affiliation{Institut f\"ur Theoretische Physik II,
  Heinrich-Heine-Universit\"at, D-40225 D\"usseldorf, Germany}
\author{Felix K\"ummel}
\affiliation{II. Institut f\"ur Physik, Universit\"at Stuttgart,
  D-70569 Stuttgart, Germany}
\author{Hartmut L\"owen}
\affiliation{Institut f\"ur Theoretische Physik II,
  Heinrich-Heine-Universit\"at, D-40225 D\"usseldorf, Germany}
\author{Clemens Bechinger}
\affiliation{II. Institut f\"ur Physik, Universit\"at Stuttgart,
  D-70569 Stuttgart, Germany}
\affiliation{Max-Planck-Institute for Intelligent Systems, D-70569 Stuttgart,
  Germany}
\author{Thomas Speck}
\affiliation{Institut f\"ur Theoretische Physik II,
  Heinrich-Heine-Universit\"at, D-40225 D\"usseldorf, Germany}

\date{\today}

\begin{abstract}
  We study experimentally and numerically a (quasi) two dimensional colloidal
  suspension of self-propelled spherical particles. The particles are
  carbon-coated Janus particles, which are propelled due to diffusiophoresis
  in a near-critical water-lutidine mixture. At low densities, we find that
  the driving stabilizes small clusters. At higher densities, the suspension
  undergoes a phase separation into large clusters and a dilute gas phase. The
  same qualitative behavior is observed in simulations of a minimal model for
  repulsive self-propelled particles lacking any alignment interactions. The
  observed behavior is rationalized in terms of a dynamical instability due to
  the self-trapping of self-propelled particles.
\end{abstract}


\maketitle


Following our physical intuition, ``agitating'' a system by, e.g., increasing
the temperature also increases disorder. The most simple and paradigmatic
example is the Ising model of interacting spins on a lattice, which, in two or
more dimensions, displays a second-order phase transition from an ordered
state to a disordered state as we increase the
temperature~\cite{chandler}. Non-equilibrium driven systems, however, may defy
our intuition and show the opposite behavior: increasing the noise strength
leads to the emergence of an ordered state~\cite{broe94,sagu07}, for example
the ``freezing by heating'' transition of oppositely driven particles in a
narrow channel~\cite{helb00}.

One class of non-equilibrium systems that currently receives considerable
attention are self-propelled, or ``active'', particles
\cite{gole05,hong07,hows07,jian10,pala10,theu12,volp11,butt12,pala13}. These
are model systems for ``living active matter'' ranging from
microtubules~\cite{sumi12} to dense bacterial
solutions~\cite{dres09,dres11,peru12} to flocks of birds~\cite{cava10}. A
common feature of many of these models is that the particle orientations
align, which leads to a multitude of collective phenomena such as
swarming~\cite{vics95} and even micro-bacterial turbulence~\cite{wens12}. This
alignment interaction can be either explicit (Vicsek-type
models~\cite{vics12}) or indirect. For example, in dense granular systems of
rods~\cite{nara07} and disks~\cite{dese10}, the combination of hard-core
repulsion and propulsion implies an effective alignment. Somewhat
surprisingly, recently it has been found that also self-propelled suspensions
lacking any alignment mechanism are able to show collective
behavior. Specifically, simulations of a minimal model for a suspension of
repulsive disks below the freezing transition~\cite{bial12} show phase
separation into a dense large cluster and a dilute gas
phase~\cite{yaou12,redn13}. Phase separation due to a density-dependent
mobility has been discussed theoretically in the context of run-and-tumble
bacteria~\cite{tail08}, and a link has been made recently to self-propelled
Brownian particles~\cite{cate13}.

Experimentally, active clustering of spherical colloidal particles has been
observed for sedimenting, platinum-coated gold particles~\cite{theu12} and
colloidal particles with an embedded hematite cube~\cite{pala13}, where
platinum and hematite act as catalysts for the decomposition of water
peroxide. In both studies, aggregation is attributed to attractive forces. In
this Letter, we investigate a suspension of carbon-coated colloidal Janus
particles dispersed in a near-critical mixture of water and lutidine. This
setup allows us to continuously vary the propulsion speed by changing the
illumination power~\cite{volp11}. We have chosen material and experimental
conditions at which the influence of attractions--due to van der Waals forces
and the phoretic motion--is largely reduced. Instead, the clustering is caused
by dynamical self-trapping of the self-propelled particles. This mechanism is
generic and does not depend on the actual means of propulsion. At higher
densities (but still below the freezing transition), we report the first
experimental data for active colloidal suspensions showing phase
separation, whereby clusters grow to a finite fraction of the system size. The
robustness of this transition is qualitatively confirmed in simulations of
purely repulsive disks.


Janus particles are prepared from SiO$_2$ beads with a radius of
$R\simeq2.13\unit{\mu m}$ by sputtering a thin layer ($10\unit{nm}$) of
graphite onto one hemisphere. These carbon-coated particles are then suspended
in a water-2,6-lutidine mixture close to the critical concentration (28 mass
\% lutidine) and a small amount of suspension is poured in a
$400\times400\unit{\mu m^2}$ cavity. The cavity is made by photolithography of
photoresist SU-8 on a glass surface. The 2D area fraction $\phi$ is tuned up
to 0.4 by adjusting the concentration of the initial suspension. The sample is
sealed with a cover glass on top. Since the height of the cavity is
about $6\unit{\mu m}$, the motion of the particles is confined to quasi-2D
although the spheres remain free to rotate in 3D.

The propulsion mechanism~\cite{volp11,butt12} is summarized as follows: The
active motion is obtained by illuminating the entire sample with a widened
laser beam with wavelength $532\unit{nm}$. The laser light is absorbed by the
carbon-coated hemisphere, which locally heats up the binary solvent above the
critical temperature. The ensuing de-mixing provides a phoretic force that
propels the particles since the two hemispheres possess different surface
properties with respect to water and lutidine. Unlike catalytic
swimmers~\cite{saba12}, e.g., particles propelled in a H$_2$O$_2$
solution~\cite{gole05,hong07,theu12,pala13} where the molecular solute is
``consumed'' by the chemical reaction, in the setup used here the environment
is not affected by the local de-mixing due to the reversibility of the
spinodal decomposition. Employed illumination intensities
($\leqslant5\unit{\mu W/\mu m^2}$) are weak compared to values reported
previously on thermophoretic motion of colloidal Janus
particles~\cite{jian10}. Moreover, particle motion is Brownian far below the
critical temperature of the water-lutidine mixture even with the illumination
turned on. Hence, we conclude that thermophoretic effects are negligible and
that diffusiophoresis is the principal propulsion mechanism~\cite{butt12}.

\begin{figure}[t]
  \centering
  \includegraphics{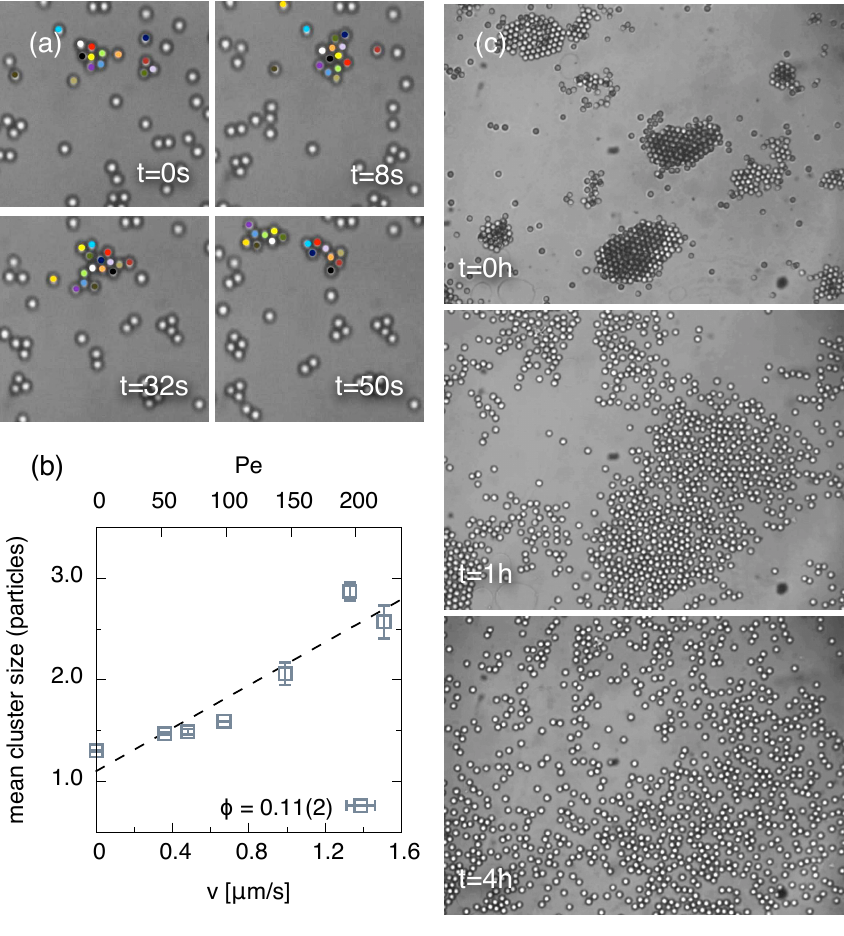}
  \caption{(a)~Dynamical clustering of self-propelled colloidal particles at
    low densities ($\phi\simeq0.1$ and $v\simeq0.65\unit{\mu m/s}$). Shown is
    the formation and breaking-up of one cluster. Every particle that at one
    time has been a member of the cluster is colored differently. (b)~The mean
    cluster size increases linearly as a function of speed $v$. The dashed
    line is the fit $1.1+1.1v$. Error bars indicate the statistical
    uncertainty. (c)~At higher densities ($\phi\simeq0.27$ and
    $v\simeq1.63\unit{\mu m/s}$), phase separation into a few big clusters and
    a dilute phase occurs. The aggregation is completely reversible: The
    snapshots show how the clusters dissolve after the illumination has been
    turned off.}
  \label{fig:clusters}
\end{figure}

The axis joining the poles of the two hemispheres defines the particle
orientation along which it is propelled. Particle motion in dilute suspensions
is described by a persistent random walk~\cite{volp11,butt12}. The measured
mean-square displacement follows the expression~\cite{hows07}
\begin{equation}
  \label{eq:msd}
  \mean{\Delta r^2} = 4D_0t + \frac{v^2}{2\Dr^2}\left[2\Dr
    t+e^{-2\Dr t}-1\right],
\end{equation}
where $D_0\simeq0.029\unit{\mu m^2/s}$ is the bare diffusion coefficient
measured in equilibrium, $v$ is the swimming speed, and $\Dr$ is the
rotational diffusion coefficient. The latter, as obtained from fits, is
independent of $v$ and approximately obeys the no-slip relation
$\Dr\approx3D_0/(2R)^2$ between translational and rotational diffusion
coefficient, showing that particles undergo free rotational diffusion. In the
following, the control parameters are the area fraction $\phi$ and the laser
intensity. In order to estimate the swimming speed $v$ in dense suspensions
for a given intensity, we determine the trajectories of isolated particles and
fit their short-time mean-square displacement to the expansion $4D_0t+(vt)^2$
of Eq.~\eqref{eq:msd}.

Under equilibrium conditions, i.e., with the illumination turned off, we
observe a homogeneous suspension at all studied area fractions
$\phi\simeq0.1$--$0.4$. After turning on the illumination, the particles are
driven out of thermal equilibrium and are propelled along their
orientation. We let the system relax into a steady state (for about 15
minutes) and then analyze trajectories with length of about 5 minutes. Typical
situations at low and high density are presented in
Fig.~\ref{fig:clusters}. At low densities, the system indeed rapidly enters a
steady state that can be described as a dynamical cluster
fluid. Fig.~\ref{fig:clusters}(a) shows the temporal evolution of a small
cluster. It clearly demonstrates that the aggregation is dynamical, i.e.,
particles join and leave the cluster, until in the last snapshot the cluster
has finally broken into two smaller clusters. Fig.~\ref{fig:clusters}(b) shows
that the mean cluster size increases approximately linearly as a function of
the propulsion speed similar to what has been observed by Theurkauff
\textit{et al.}~\cite{theu12}.

At higher densities, we observe a phase separation
[cf. Fig.~\ref{fig:clusters}(c)] where clusters grow until the system consists
of a few big clusters surrounded by a dilute gas phase. We presume that the
final stage is the condensation into one cluster. However, the slow dynamics
of the large aggregates puts the direct observation of this final stage out of
our current reach. Fig.~\ref{fig:clusters}(c) shows the temporal evolution of
the sample after we turn off the illumination. Particle diffusion restores the
homogeneous density profile, indicating that also for large clusters the
aggregation is reversible and solely induced by the propulsion of the
colloidal particles.


We employ computer simulations of a minimal model~\cite{bial12,yaou12,redn13}
both in order to corroborate our experimental conclusions and to test the
limits of simplified mathematical models applied to self-propelled
suspensions. The model is defined as follows: We simulate $N=4900$ interacting
particles, each of which has an orientation $\vec e_k$ diffusing freely about
the $z$-axis with rotational diffusion coefficient $\Dr$. Particles move in
two dimensions in a quadratic box with periodic boundary conditions. In
addition to translational Brownian motion, particles are propelled along their
orientation with a constant speed. Moreover, we neglect hydrodynamic
interactions between colloidal particles. The coupled equations of motion are
\begin{equation}
  \label{eq:model}
  \dot\x_k = -\nabla_k U + \Pe\;\vec e_k + \nois_k
\end{equation}
for the particle positions $\{\x_k\}$, where the Gaussian white noise
$\nois_k$ models the coupling to the solvent and
$U=\sum_{k<k'}u(|\x_k-\x_{k'}|)$ is the potential energy due to pairwise
interactions. Quantities are made dimensionless using $2R$ as unit of length
and $(2R)^2/D_0$ as unit of time, which implies the P\'eclet number
$\Pe=2Rv/D_0$. The equations of motion~\eqref{eq:model} are integrated with a
(minimal) time step $10^{-5}$.

\begin{figure}[t]
  \centering
  \includegraphics{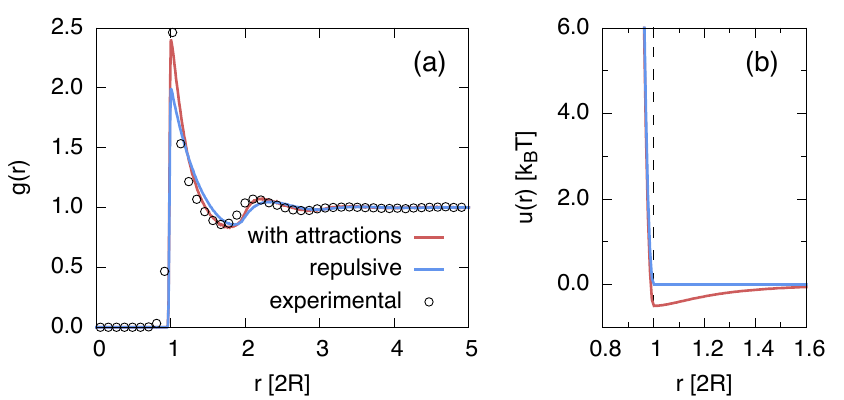}
  \caption{(a)~Structure of the passive suspension (without illumination):
    Experimental pair distribution function $g(r)$ at packing fraction
    $\phi\simeq0.37$ and simulation results employing the potential
    Eq.~\eqref{eq:pot} for $\lam=0$ (repulsive) and $\lam=0.5\kT$ (slight
    attractions). (b)~Corresponding pair potentials.}
  \label{fig:wca}
\end{figure}

The experiments are carried out in a quasi two-dimensional geometry. Particles
may move out of plane and slightly overlap in the recorded images. To account
for this (apparent) softness in the simulations, for the pair interactions we
choose
\begin{equation}
  \label{eq:pot}
  u(r) =
  \begin{cases}
    \eps u_\text{LJ}(r) + u_\text{LJ}(2R)(\lam-\eps) & (r\leqslant 2R) \\
    \lam u_\text{LJ}(r) & (r>2R)
  \end{cases}
\end{equation}
with Lennard-Jones potential
$u_\text{LJ}(r)=4\left[(\sig/r)^{12}-(\sig/r)^6\right]$. That is, we use a
repulsive core (the WCA potential) to which optionally we add an attractive
tail of depth $\lam$~\cite{week72}. For the passive equilibrium system (no
illumination), Fig.~\ref{fig:wca}(a) compares the experimentally measured
radial distribution function $g(r)$ with the simulation results, both at area
fraction $\phi\simeq0.37$. We fix $\eps=100\kT$ and
$\sig/(2R)=2^{-1/6}\simeq0.891$ such that the potential minimum coincides with
the particle diameter, see Fig.~\ref{fig:wca}(b). Good agreement between
experiment and simulations is achieved by adding a small attraction with
$\lam=0.5\kT$ in the simulations. However, in the following we will focus on
the purely repulsive pair potential with $\lam=0$ to show that, conceptually,
the observed phenomena do not depend on attractions.

\begin{figure}[t]
  \centering
  \includegraphics{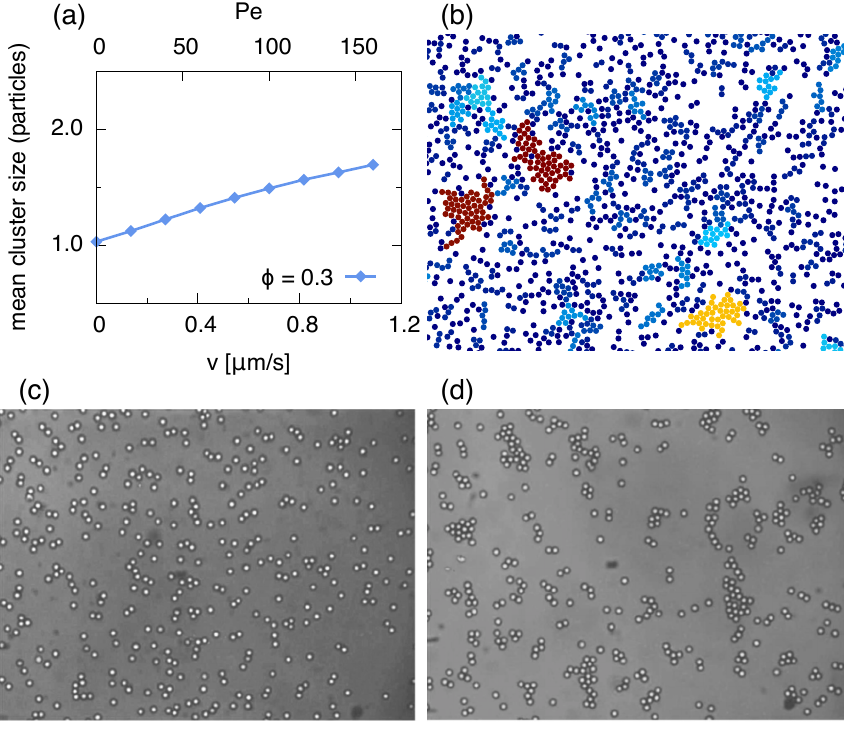}
  \caption{Simulation results for the clustering of purely repulsive discs at
    area fraction $\phi=0.3$: (a)~Mean cluster size as a function of swimming
    speed $v$. (b)~Snapshot for speed $\Pe=140$ (corresponding to
    $v\simeq0.95\unit{\mu m/s}$). Clusters with same relative size have the
    same color. For comparison: snapshots (c) and (d)~show the experimental
    suspension for speeds (c)~$v\simeq0.36\unit{\mu m/s}$ and
    (d)~$v\simeq1.51\unit{\mu m/s}$.}
  \label{fig:low}
\end{figure}

Clusters are determined from a simple overlap criterion: In the simulations,
all particles with a separation smaller than their diameter share a
``bond''. A cluster is then the set of all particles that are mutually
bonded. For the experimental trajectories we use a slightly different method,
where we estimate cluster sizes through the enclosed area since within larger
clusters it becomes difficult to reliably detect particle positions. The
measured mean cluster size in Fig.~\ref{fig:clusters}(b) increases linearly as
a function of the speed, i.e., the driving stabilizes small clusters. As shown
in Fig.~\ref{fig:clusters}(a), these clusters are dynamical and not the result
of irreversible aggregation. The simulation results shown in
Fig.~\ref{fig:low}(a) demonstrate that a purely repulsive pair potential is
sufficient to reproduce the increase of the mean cluster size with swimming
speed. However, comparing with Fig.~\ref{fig:clusters}(b), the increase of the
cluster size is somewhat stronger in the experiments. The snapshots of the
simulations [Fig.~\ref{fig:low}(b)] and experiments [Fig.~\ref{fig:low}(c) and
(d)] reveal another difference: while in the simulations a few large clusters
dominate, the experimental snapshots show many clusters containing fewer
particles. These differences are most likely due to the influence of
hydrodynamics, see also Supplementary Material for more details. Hydrodynamic
aggregation of swimmers has been demonstrated in simulations~\cite{ishi08a}
and experimentally~\cite{dres09,thut11}.

\begin{figure}[t]
  \centering
  \includegraphics{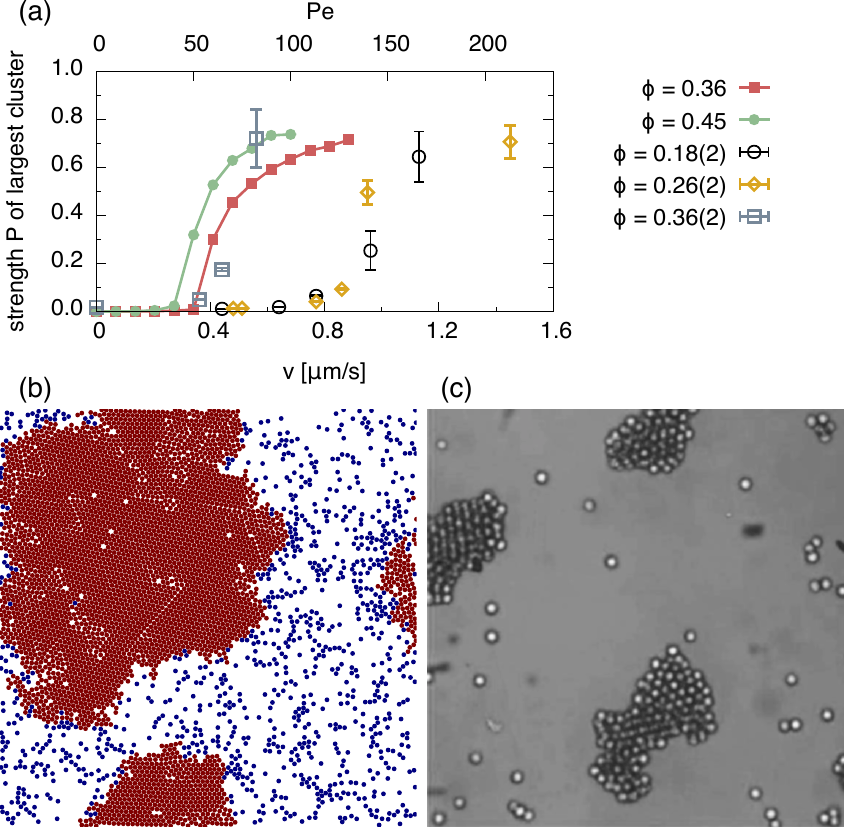}
  \caption{Phase separation: (a)~Relative mean size $P$ of the largest cluster
    as a function of swimming speed $v$. Shown are experimental results (open
    symbols) and simulation results (closed symbols). (b)~Simulation snapshot
    of the separated system at $\phi=0.5$ and speed
    $\Pe=100$. (c)~Experimental snapshot at $\phi\simeq0.25$ and
    $v\simeq1.45\unit{\mu m/s}$.}
  \label{fig:phasesep}
\end{figure}

Increasing the density, we observe a transition from the initially disordered,
homogeneous fluid into an ordered state as we change the swimming speed, see
Fig.~\ref{fig:clusters}(c) and Fig.~\ref{fig:phasesep}(c). The ordered state
is comprised of a few big clusters surrounded by a dilute phase of single
self-propelled particles. In the simulations, clusters finally merge into a
single big cluster. This cluster is not static but it is constantly changing
its shape while particles are exchanged between the cluster and the diluted
phase.

As a geometrical order parameter for the transition, we use the average
fraction $P=\mean{N_\text{lc}}/N$ of particles in the largest cluster. In one
configuration, $N_\text{lc}$ is the number of particles that are part of the
largest cluster. For the experimental data, we actually add together the size
of all clusters larger than $N/10$ particles since we expect all big clusters
to finally merge. We only observe the coalescence of smaller cluster and not
that a larger cluster breaks up. The order parameter is plotted in
Fig.~\ref{fig:phasesep}(a) as a function of the swimming speed $v$. At some
critical speed it shows a transition from the disordered fluid into the
ordered phase, wherein the largest cluster occupies a finite fraction of the
system. The ordered state is thus reached by \emph{increasing} the driving
strength. The critical speed is shifted to lower values at higher
densities. The transition occurs in the experiments already at densities that
are lower than what is predicted in the simulations. For the highest
experimental density $\phi\simeq0.36$, the critical speed agrees quite well
with the simulations.


What is the mechanism of the cluster formation? Of course, clusters also form
in equilibrium systems if attractions between particles are present. For large
enough attraction, the gain of energy overcomes the loss of entropy and the
suspension separates into a dense liquid or solid, and a dilute gas
phase. Thermodynamically stable cluster fluids generally require long-ranged
repulsion (e.g., charged particles) together with short-ranged
attraction~\cite{scio04} (however, stable cluster fluids in colloidal
suspensions have been reported also in the absence of long-ranged
repulsion~\cite{lu07}). Because in our experiments we have used carbon instead
of a metal as coating material for the Janus particles, we have largely
reduced attraction-driven aggregation of particles due to short-ranged van der
Waals forces~\cite{israel}. Moreover, phoretic attractions as well as
alignment interactions can be neglected for the experimental conditions used,
see Supplementary Material.

\begin{figure}[t]
  \centering
  \includegraphics{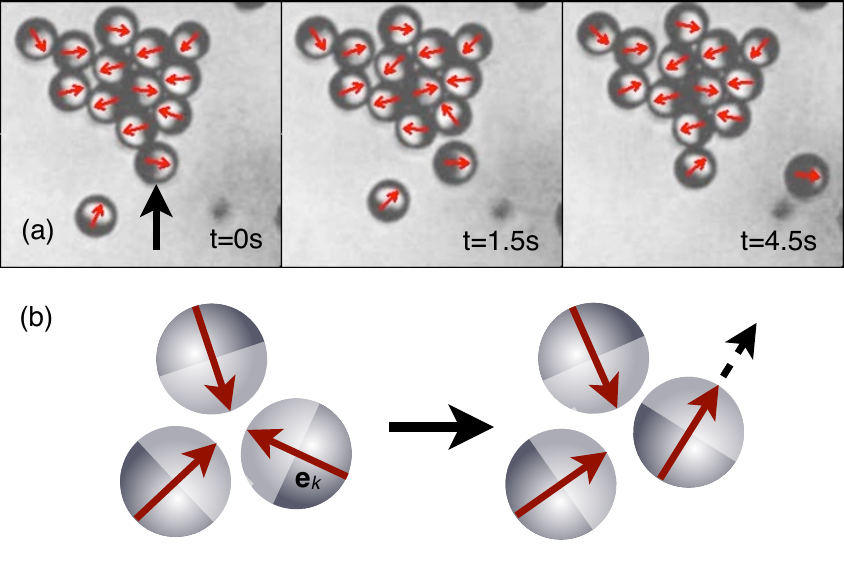}
  \caption{(a)~Consecutive close-ups of a cluster, where we resolve the
    projected orientations (arrows) of the caps. Particles along the rim
    mostly point inwards. The snapshots show how the indicated particle
    towards the bottom (left) leaves the cluster (center) and is replaced by
    another particle (right). (b)~Sketch of the self-trapping mechanism: for
    colliding particles to become free, they have to wait for their
    orientations to change due to rotational diffusion and to point outwards.}
  \label{fig:trap}
\end{figure}

To further investigate the clustering mechanism, we have repeated the
experiments using larger particles with radius $R\simeq4\unit{\mu m}$, which
allow us to resolve the caps and thus the projected orientations of particles
(dynamics is also much slower, which is why for measurements we have employed
smaller particles). Fig.~\ref{fig:trap}(a) shows consecutive snapshots of a
single cluster. Note that the orientations along the rim mostly point
inwards. One particle with an outward orientation leaves the cluster while
another particle attaches. The emerging physical picture is thus that of a
simple self-trapping mechanism, see Fig.~\ref{fig:trap}(b): Two or more
particles that collide head-on are blocked due to the persistence of their
orientations. Hence, a particle situated in the rim of the cluster has to wait
a time $\sim1/\Dr$ until rotational diffusion points its orientation outward
to become free again. While the time to leave the cluster is independent of
the swimming speed $v$, a larger swimming speed implies a larger probability
for other particles to collide with the cluster, leading to its growth. The
size of clusters is determined by the flux balance of incoming and outgoing
particles.


To summarize, we have presented experimental results for a colloidal
suspension of Janus particles that are self-propelled through the heating of a
carbon-coated hemisphere in a near-critical binary mixture of water and
lutidine. At low densities, we observe the emergence of dynamical
clusters. The mean cluster size increases approximately linearly with the
propulsion speed in agreement with previous work using catalytic
swimmers~\cite{theu12}. At higher densities, the suspension separates into big
clusters surrounded by a dilute phase of free swimmers. Both phenomena are
captured qualitatively by Brownian dynamics simulations of a minimal model
without any alignment interaction and neglecting hydrodynamics.


We thank Giovanni Volpe, Matthieu Marechal, and Marco Heinen for important
discussions. This work has been supported financially by the DFG within SFB
TR6 (project C3). IB is supported by the Marie Curie ITN Comploids, funded by
the EU Seventh Framework Program (FP7).



\newpage

\begin{figure*}[h]
  \centering
  \includegraphics{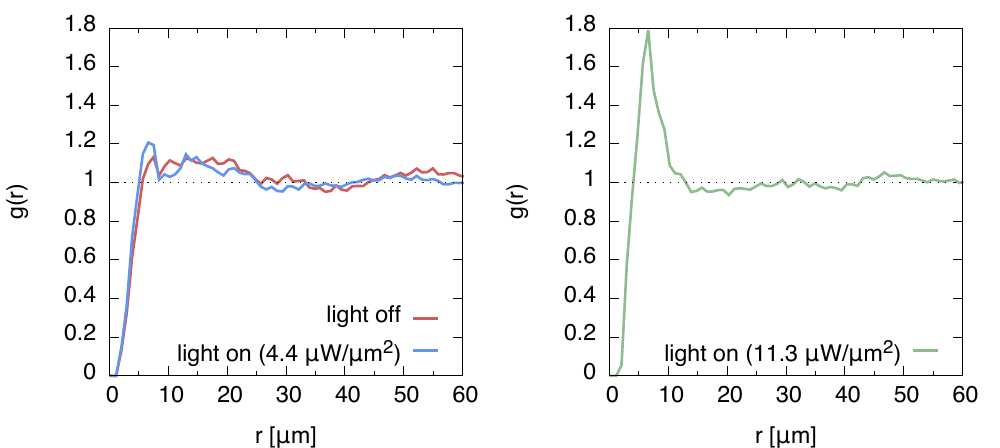}
  \caption{Pair distribution function of passive tracer particles in the
    presence of an active coated Janus particle fixed at the origin: (left)
    Without illumination and at moderate illumination intensity $4.4\unit{\mu
      W/\mu m^2}$. (right) At high illumination intensity $11.3\unit{\mu W/\mu
      m^2}$.}
  \label{fig:phoresis}
\end{figure*}

\begin{figure*}[t]
  \centering
  \includegraphics{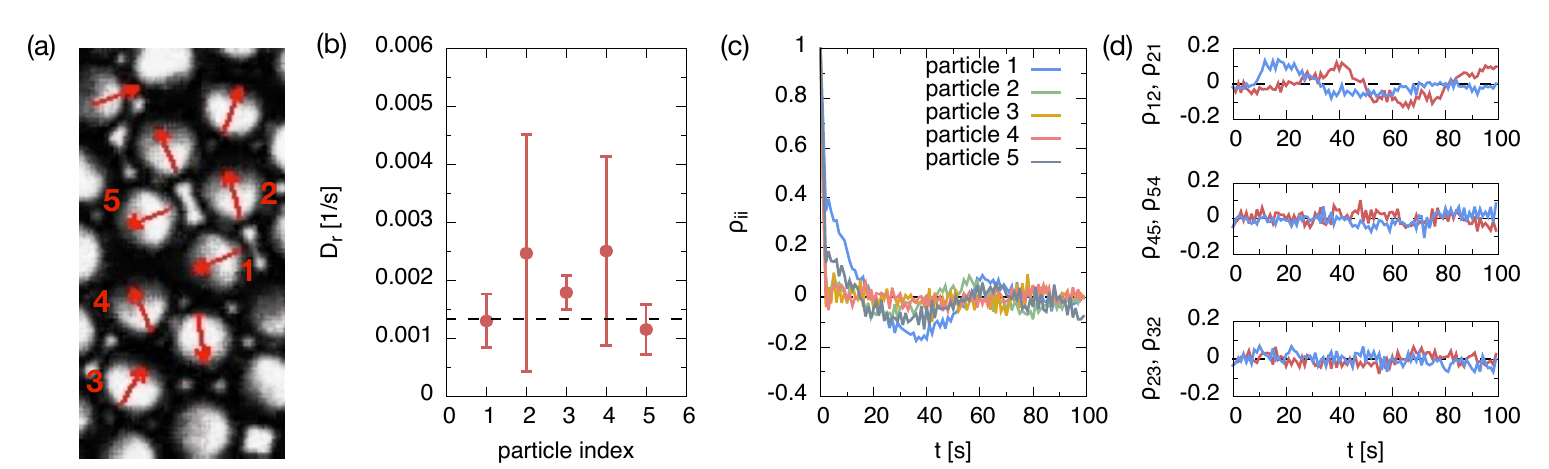}
  \caption{(a)~Snapshot of the colloidal suspension. The measured projected
    orientations are marked by red arrows. The index of the particles that we
    have analyzed is labeled. (b)~Estimated apparent rotational diffusion
    coefficient as a function of the particle index. The dashed line shows the
    estimated diffusion coefficient $3D_\text{b}/(2R)^2$ of a free
    particle. (c)~Normalized auto-correlations $\rho_{ii}(t)$ as a function of
    time difference $t$. (d)~Off-diagonal elements of $\rho_{ij}(t)$ for three
    particle pairs.}
  \label{fig:align}
\end{figure*}

\begin{figure*}[t]
  \centering
  \includegraphics{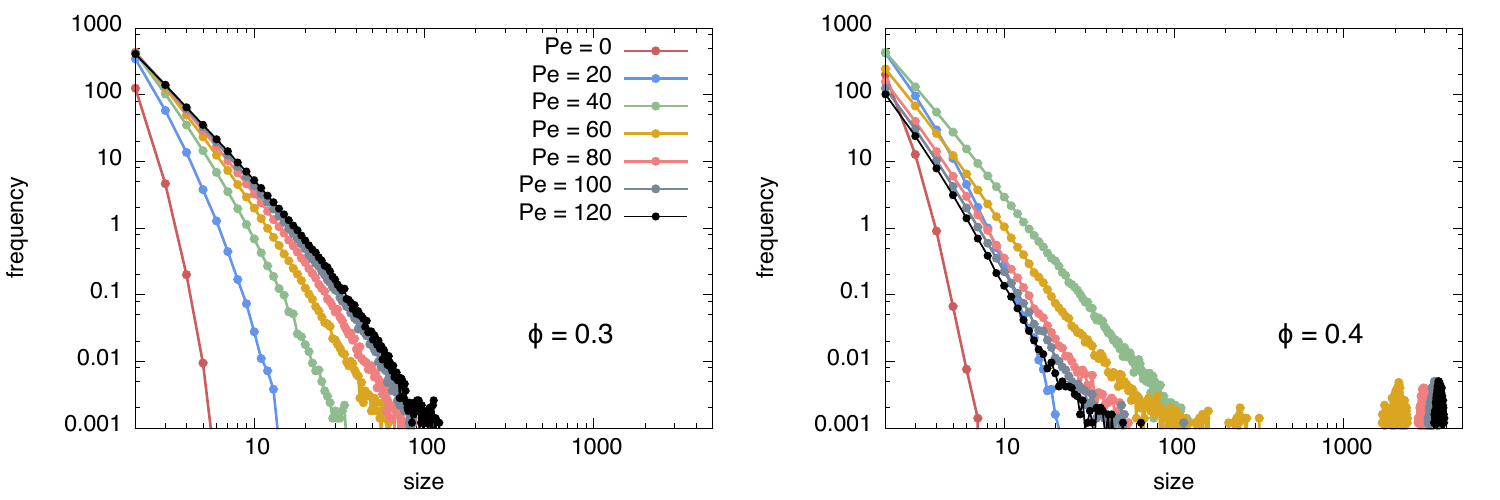}
  \caption{Cluster size distributions for $\phi=0.3$ (left) and $\phi=0.4$
    (right).}
  \label{fig:dist}
\end{figure*}

\section{Supplementary Information}


\subsection{Phoretic attractions}

We have performed an experiment similar in spirit to what has been done in
Ref.~\citenum{pala13} in order to address a possible phoretic attraction
between the self-propelled colloidal particles. To this end we have prepared a
cell containing a suspension of passive tracer particles with radius
$R\simeq1.3\unit{\mu m}$ and a few coated Janus particles with radius
$R\simeq4\unit{\mu m}$. The volume fraction of the passive particles is
$\phi\simeq0.3$.

We have picked a region of the sample where by chance one of the coated
particles sticks to the surface. To study whether attractive forces between
active and passive particles occur, we have calculated the pair distribution
function $g(r)$ of the passive particles with the active coated particle at
the origin. The result is shown in Fig.~\ref{fig:phoresis}. We have
investigated three situations: illumination turned off, illumination with
moderate intensity $4.4\unit{\mu W/\mu m^2}$, and with high intensity
$11.3\unit{\mu W/\mu m^2}$. The swimming speeds measured using the method
described in the main text are $v\simeq1.4\unit{\mu m/s}$ and
$v\simeq9\unit{\mu m/s}$, respectively. Note that the maximal intensity used
to gather the data presented in the main text is $5\unit{\mu W/\mu m^2}$.

At high laser power we do indeed observe an aggregation of passive particles
around the immobile coated particle indicating an effective phoretic
attraction. However, at the lower laser power corresponding to the actual
experiments we do not observe aggregation. We thus conclude, at least for the
illumination intensities used, that phoretic attractive forces can be
neglected and that the observed phase separation is due to the self-trapping
of active particles.


\subsection{Alignment of orientations}

To address a possible alignment of orientations we have prepared a dense
sample of active particles with radius $R\simeq3.88\unit{\mu m}$ at
illumination intensity $1.88\unit{\mu W/\mu m^2}$. In Fig.~\ref{fig:align}(a)
a snapshot of the sample is shown. The particle orientations appear to be
random without any alignment. For a more quantitative analysis we have
recorded the projected angle $\vhi_i(t)$ with a time resolution of $1\unit{s}$
for the labeled particles. From the time series we estimate the angular
velocity
\begin{equation*}
  \dot\vhi_i(t) \simeq \vhi_i(t+1)-\vhi_i(t).
\end{equation*}
The correlations as a function of time difference are given by
\begin{equation*}
  \sig^2_{ij}(t) = \frac{1}{K}\sum_{k=0}^{K-1} \dot\vhi_i(k)\dot\vhi_j(k+t).
\end{equation*}
From these correlations we can extract a rough estimate
$\Dr\simeq\frac{1}{2}\sig^2_{ii}(0)$ for the rotational diffusion coefficient
for particle $i$, which is shown in Fig.~\ref{fig:align}(b). Error bars are
estimated as the variance when splitting the data into three sets. Also shown
is the rotational diffusion coefficient for a free particle
$\Dr=3D_\text{b}/(2R)^2\simeq1.3\times10^{-3}\unit{1/s}$ employing the no-slip
boundary condition as appropriate for colloidal particles. Here,
\begin{equation*}
  D_\text{b} = \frac{kT}{6\pi\eta R} \simeq 0.027 \unit{\mu m^2/s}
\end{equation*}
is the bare translational diffusion coefficient with $T\simeq30^\circ$C and
the viscosity of water-lutidine
$\eta\simeq2.1\times10^{-3}\unit{kg/(ms)}$~\cite{stei72}. We find that the
estimated rotational diffusion coefficients are reasonably close to that of a
free particle.

Particles 2 and 4 seem to rotate faster but also their statistical error is
much larger. For an explanation note that we are only able to measure the
\emph{projected} angle $\vhi$ whereas the particle rotates in three dimensions
described by spherical coordinates $\vhi$ and $\theta$. For free rotation the
stochastic equations of motion read~\cite{raib04}
\begin{equation}
  \label{eq:vhi}
  \dot\vhi = \frac{\xi_\vhi}{\sin\theta}, \qquad
  \mean{\xi_\vhi(t)\xi_\vhi(t')} = 2\Dr\delta(t-t'),
\end{equation}
and
\begin{equation}
  \dot\theta = \frac{1}{\tan\theta} + \xi_\theta, \qquad
  \mean{\xi_\theta(t)\xi_\theta(t')} = 2\Dr\delta(t-t').
\end{equation}
Our estimate for $\Dr$ is only accurate for $\theta\simeq\pi/2$, i.e., the
orientation of the particle is parallel to the top and bottom slides of the
cell. If the orientation of the particle has moved out of this plane it will
appear to rotate faster, which explains the data for particles 2 and 4.

The normalized correlation coefficients between particles $i$ and $j$ read
\begin{equation*}
  \rho_{ij}(t) = \frac{\sig^2_{ij}(t)}{\sig_{ii}(0)\sig_{jj}(0)}.
\end{equation*}
The auto-correlations $\rho_{ii}(t)$ are plotted in
Fig.~\ref{fig:align}(c). In agreement with Eq.~\eqref{eq:vhi}, most of the
curves decay very fast from $\rho_{ii}(0)=1$ to zero, i.e., kicks are
uncorrelated in time. However, particles 1 and 5 show some oscillations
indicating a memory. This memory might arise from hydrodynamic interactions
with other particles or with the substrate. In Fig.~\ref{fig:align}(d) we plot
the time-dependent off-diagonal elements for three particles pairs. For
particles 2 and 3 (bottom panel) as well as 4 and 5 (center panel), which are
neighbors but with a larger separation, no correlations are found. Moreover,
$\rho_{ij}(0)\simeq0$ for all particle pairs. There are, however, systematic
oscillations for the neighboring particles 1 and 2 (top panel) in response to
an earlier motion of the other particle.

The picture that emerges from this data is that there is no systematic
alignment of particle orientations, which justifies their neglecting in the
minimal model. However, some interactions are present especially at small
particle separations.


\subsection{Cluster size distributions}

For completeness, in Fig.~\ref{fig:dist} we show numerical results for the
distribution of cluster sizes for two densities and different swimming speeds
$\Pe$. For the homogeneous fluid at $\phi=0.3$ clusters become larger as we
increase the swimming speed. For the higher density the same is observed up to
the critical speed $\simeq 40$, where the distribution acquires a power
law. Such a change to a power law has also been observed in a dense bacterial
colony~\cite{peru12}. Going beyond the critical speed, phase separation sets
in which is reflected in the distribution by a second hump at large size
corresponding to the largest cluster fluctuating in size. Clusters in the
dilute phase are less frequent as we increase the speed since the size of the
largest cluster also increases.

\end{document}